\documentclass[prl, twocolumn, superscriptaddress]{revtex4-1}
\usepackage{graphicx}
\usepackage{latexsym}
\usepackage{amsmath}
\usepackage{amsthm}
\usepackage{amssymb}
\usepackage{epstopdf} 
\usepackage{enumerate}
\usepackage{setspace}   
\usepackage{dcolumn}
\usepackage{bm}
\usepackage{setspace}   
\usepackage{slashed}

\begin{document}
\title{Non-Fermi liquid and topological states with strong spin-orbit coupling } 

\author{Eun-Gook Moon }
\affiliation{Department of Physics, University of California, Santa Barbara, CA 93106, USA}

\author{Cenke Xu }
\affiliation{Department of Physics, University of California, Santa Barbara, CA 93106, USA}

\author{Yong Baek Kim }
\affiliation{Department of Physics, University of Toronto, Toronto, Ontario M5S 1A7, Canada}

\author{Leon Balents }
\affiliation{Kavli Institute for Theoretical Physics, University of California, Santa Barbara, CA 93106, USA}

\date{\today}

\begin{abstract}
  We argue that a class of strongly spin-orbit coupled materials,
  including some pyrochlore iridates and the inverted band gap
  semiconductor HgTe, may be described by a minimal model consisting
  of the Luttinger Hamiltonian supplemented by Coulomb interactions, a
  problem studied by Abrikosov and collaborators.  It contains two-fold
  degenerate conduction and valence bands touching quadratically at
  the zone center. Using modern renormalization group methods, we
  update and extend Abrikosov's classic work and
  show that interactions induce a quantum critical non-Fermi liquid
  phase, stable provided time-reversal and cubic symmetries are
  maintained.  We determine the universal power-law exponents
  describing various observables in this ``Luttinger Abrikosov
  Beneslavskii'' state, which include conductivity, specific heat,
  non-linear susceptibility and magnetic Gruneisen number.
  Furthermore, we determine the phase diagram in the presence of cubic
  and/or time-reversal symmetry breaking perturbations, which includes
  topological insulator and Weyl semi-metal phases.  Many of these
  phases possess an extraordinarily large anomalous Hall effect, with
  the Hall conductivity scaling {\sl sub-linearly} with magnetization,
  $\sigma_{xy}\sim M^{0.51}$.  
\end{abstract}

\maketitle

Divining the nature of critical non-Fermi liquid (NFL) phases of
electrons is one of the most outstanding problems in correlated
electron systems, such as cuprates, pnictides, and heavy fermion
materials \cite{louis,hirsch,chubukov,gegenwart}.  Theory, spurred
largely by NFL behavior in the cuprates, has focused on models with
large Fermi surfaces, appropriate to these materials.  Though there
have been promising technical advances\cite{max}, this approach has
not provided clear resolution of experimental puzzles.  Recent theory and
experiment have uncovered a new frontier for correlated phenomena:
Mott correlation physics in materials with strong spin-orbit coupling
(SOC) \cite{pesin,bjkim1}. Of particular interest in this regard are
the 5d transition metal oxides, where compelling evidence has been
built for Mott phenomena driven cooperatively by SOC and Coulomb
interactions \cite{pesin,bjkim1,takagi}.  Here we consider the possibility of
distinct NFL states in the strong SOC regime. 

We are particularly motivated by the pyrochlore iridates, A$_2$Ir$_2$O$_7$, where A
is a lanthanide element \cite{maeno,matsuhira}. These materials display $T>0$
metal-insulator transitions, with the metal-insulator transition
temperature decreasing with increasing A radius.  The progression
culminates with Pr$_2$Ir$_2$O$_7$, which is a highly unconventional
metal down to the lowest temperatures \cite{nakatsuji,machida}. 
It displays logarithmic NFL behavior of
the magnetic susceptibility, and a remarkable enormous zero field
anomalous Hall effect (AHE), in the absence of any measurable
magnetization \cite{nakatsuji,machida}.  

With these studies as motivation, we utilize prior studies of the
electronic structure of pyrochlore iridates\cite{ashvin,ybkim1,ybkim2} to
show that a minimal description for the electronic states in their
paramagnetic phase is a storied Hamiltonian from semiconductor
physics: the Luttinger model of inverted band gap semiconductors \cite
{luttinger}. This model has gained recent notoriety for its relevance
to HgTe, the starting material for some topological insulators
\cite{TI,qi,zhang,zhang2}.  While HgTe is a weakly correlated material
where band structure alone provides a sufficient description, in the
5d materials the Luttinger Hamiltonian must be supplanted by
interactions.

In this paper, we carry out such an analysis, rediscovering and
extending a storied analysis by Abrikosov and Beneslavskii of Coulomb
forces on the Luttinger problem \cite{ab1,ab2}. Using modern renormalization group (RG)
techniques, we confirm Abrikosov's conclusion that long-range Coulomb
interactions convert the quadratic band touching into a quantum
critical NFL, prove the stability of the state within an $\epsilon$
expansion, and calculate the full set of anomalous dimensions
characterizing the state. Consequently, we call the resulting phase a
{\em Luttinger-Abrikosov-Beneslavskii} (LAB) state.  While the LAB
phase is stable in the presence of time-reversal and cubic symmetries,
we show that it is a ``parent'' state for other exotic states that can
be reached by breaking either or both of these: metallic and 
double-Weyl semi-metallic phases with enormous anomalous Hall effects
(AHEs), and topological insulators.  
We discuss the implications for the iridates
at the end of the paper.

The LAB phase itself has striking properties.  Its NFL nature is
revealed directly by algebraic singularities in the electron spectral
function (probed in angle-resolved photoemission) and in optical
conductivity, as well as indirectly through many power law
thermodynamic and response functions.  

\begin{figure}
\includegraphics[width=3.3in]{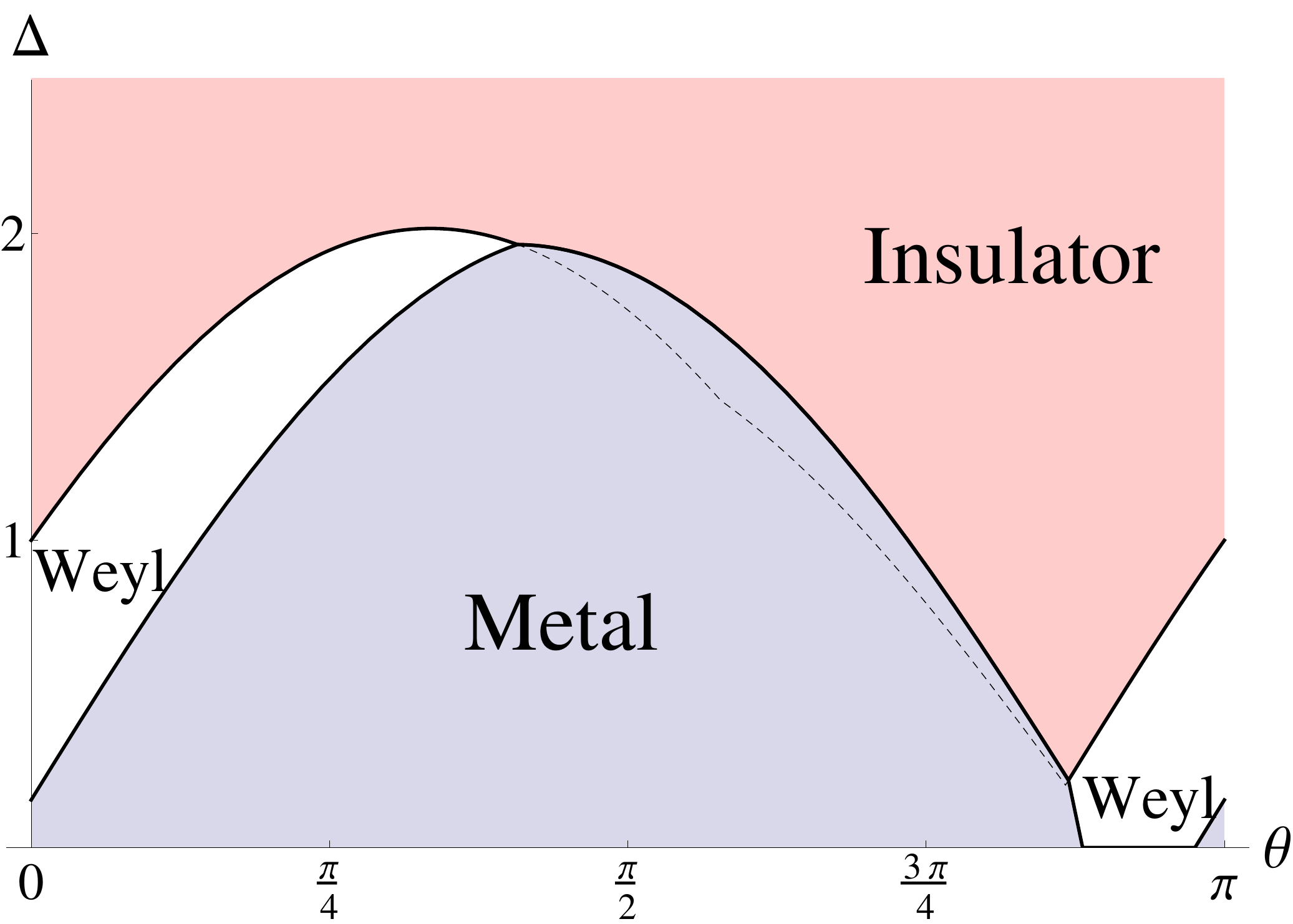}
\caption{ Phase diagram of the perturbed LAB in the space of
  renormalized strain to Zeeman field ratio, $\Delta \equiv
  (\delta/H)_R$, versus cubic Zeeman angle, $\theta$. Here``Weyl''
  denotes the (double) Weyl semimetal, ``Ins.''  insulator, and
  ``Metal'' a metallic phase which has Weyl points shifted from the
  Fermi energy in the region below the dashed line. For $H=0$, the insulator is
  a topological insulator.  } \label{fig1}
\end{figure}

We now turn to the exposition of these results.  We consider the {\sl
  paramagnetic} band structure based on prior work
\cite{ashvin,ybkim1,ybkim2}, which argued that the states at the zone
center ($\Gamma$ point) near the Fermi energy are comprised of the
four-dimensional representation, which can be described by ``angular
momentum'' operators $\vec{J}$ (which are $j=3/2$ matrices)
transforming as the T$_2$ representation of the cubic group.  In our
minimal model, we assume only these states close to $\Gamma$ are
important.  Then $k\cdot p$ theory and cubic symmetry determines the
band structure in its vicinity to be precisely described by the
Luttinger Hamiltonian with three effective mass parameters
\cite{luttinger,murakami},
\begin{eqnarray}
\mathcal{H}_0(k)&=&\frac{k^2}{2 \tilde{M}_0} + \frac{ \frac{5}{4}k^2-(\vec{k} \cdot \vec{J})^2}{2m} - \frac{ (k_x^2 J_x^2 + k_y^2 J_y^2 +k_z^2 J_z^2)}{2 M_c} .\nonumber
\end{eqnarray}
This describes doubly degenerate bands with energy
\begin{eqnarray}
E_{\pm}(k)= \frac{k^2}{2M_0} \pm \sqrt{\left(\frac{k^2}{2m}\right)^2+ \frac{m+2 M_c}{4 m M_c^2} p_c(k) },
\end{eqnarray}
where $ p_c(k) = \sum_{i} k_i^4 -\sum_{i \neq j} k_i^2 k_j^2$ and 
 $M_0 = (4 M_c \tilde{M}_0)/(4M_c-5 \tilde{M}_0)$. 
Henceforth we assume $M_0 > m$, which describes conduction and valence
bands touching quadratically at $E=k=0$, where the chemical potential for the 
undoped material crosses.

The LAB is obtained by adding to this the long-range Coulomb
interaction. We implement the latter
by a scalar potential $\varphi$, which in the Euclidean path integral
formalism gives the action
\begin{eqnarray}
&& {S}_{L} = \int d \tau d^d x \left\{ \psi^{\dagger}  \Big[\partial_{\tau}
-i e \, \varphi +\hat{\mathcal{H}}_0\Big] \psi + \frac{c_0}{2} (\partial_i
\varphi)^2 \right\}\!, \label{eq:2}
\end{eqnarray}
with $\hat{\mathcal{H}}_0 = \mathcal{H}_0(-i\vec\nabla)$ and
$c_0=1/4\pi$. Here $\psi$ is a four-component spinor, but subsequently we
will artificially add an additional $U(N_f)$ flavor index, which
allows a check on our calculations by large $N_f$ methods; the
physical case is $N_f=1$.
Eq.~\eqref{eq:2} contains in addition to the three mass parameters,
the Coulomb coupling constant $e$.  For $e=0$, scale invariance is
manifest, with the scaling dimensions $[x^{-1}]=1 \, , \, [\tau^{-1}]=z
\, , \, [\psi] = \frac{d}{2} \, , \, [\frac{1}{m}]=z-2$, $[\varphi]=(d+z-2)/2$. Here we
introduce the dynamic critical exponent ($z$), which is naturally
$z=2$ with $e=0$, but will become non-trivial with interactions.

Directly in the physical case $d=3$, the dimension of the coupling
constant is $[e^2] =1$, so Coulomb interactions are strongly relevant.
Therefore we employ the $\varepsilon=4-d$ expansion to control the RG
analysis.  As familiar from quantum electrodynamics, three one loop
Feynmann diagrams contribute to leading order in $\varepsilon$: the
fermion self-energy, boson self-energy, and vertex correction.  Here
we show that the relevance of Coulomb interactions signals, rather
than a flow to strong coupling and a symmetry breaking instability,
the formation of a new {\em stable} interacting fixed point, 
which describes the critical non-Fermi liquid LAB state (Abrikosov's
analysis tacitly assumes this stability).

The RG is carried out perturbatively in $e$, but non-perturbatively in
the mass parameters.  Thus a full treatment gives non-trivial and
complete beta functions for the two dimensionless mass ratios $m/M_0$,
$m/M_c$; these are given in the Supplementary Material.  The
analysis of the full RG shows, however, that there is a single stable
{\sl isotropic} fixed point corresponding to $m/M_0=m/M_c=0$, so for simplicity we
quote in the main text only the results in the vicinity of this point.  

In this limit, the leading contribution to the bosonic self energy becomes
\begin{eqnarray}
\frac{1}{N_f}\Sigma_{\varphi}(q, 0) 
&=& -(2m) e^2 \left[ \int \! \frac{d^dk}{(2\pi)^d}\, \frac{1}{k^4}\right] \times q^2 ,
\end{eqnarray}
where we took the $\omega \rightarrow 0$ limit because frequency
dependence is subdominant.  The divergence should be absorbed by
rescaling the bosonic field, $\varphi \rightarrow e^{-\eta_b
  d\ell}\varphi$ upon reduction of the hard momentum cutoff $\Lambda
\rightarrow e^{-d\ell}\Lambda$, which defines the RG parameter $\ell$.
This gives the bosonic anomalous dimension $\eta_b = 2
  N_f  u$ \cite{large_n}, where the dimensionless coupling
constant is $u= \frac{m\, e^2 }{ 8\pi^2 c_0 \Lambda^{4-d}}$, which has the
physical meaning in $d=3$ of the ratio of the real space cutoff to the
effective Bohr radius.  
The frequency dependence of the one loop fermionic self-energy and the vertex
correction both vanish, the result of a Ward
identity. 
For $k\neq 0$, the fermion self-energy gives mass corrections,
e.g. $\delta (1/m) = 8u/(15m) \times d\ell$ to leading order.
Detailed analysis is given in the Supplementary Material. 

Given these calculations, we choose $z=2- 8u/15$ to keep the
mass $m$ fixed, which gives the RG equations, to lowest order in
$m/M_c, m/M_0$:
\begin{eqnarray} \label{betaftn} 
&& \frac{d }{d \ell} u = \varepsilon \, u - \frac{30N_f+8}{15}
u^2, \\ 
&&  \frac{d}{d \ell}\left(\tfrac{m}{M_c}\right ) = -0.152 u
 \left(\tfrac{m}{M_c}\right) ,  \frac{d}{d \ell}\left(\tfrac{m}{M_0}\right ) = - \tfrac{8}{15}u 
 \left(\tfrac{m}{M_0}\right)  .\nonumber 
\end{eqnarray}

From the first equation above we find the fixed point coupling and
hence dynamical exponent,
\begin{equation}
u^* =  \frac{15}{30 N_f+8 } \varepsilon  
\, ,\,  z=2-\frac{4}{15 N_f+4 } \varepsilon ,
\end{equation}
and since $u^*>0$, the second line in Eq.~\eqref{betaftn} implies both
$m/M_0$ and $m/M_c$ are irrelevant. This establishes the existence and
nature of the stable, isotropic fixed point describing the LAB phase.
 As a check, we have carried out a large $N_f$ expansion, which gives the same bosonic 
anomalous dimension as in the $\varepsilon$ expansion at the one-loop
level, supporting the stability of the LAB phase. 

The presence of the stable interacting fixed point can be understood
physically as a balance of partial dynamical screening of the Coulomb
interactions by electron-hole pairs and mass enhancement of the same
quasiparticles by pairs.   This situation is in sharp contrast to the case of a vanishing {\em
  indirect} band gap, for which to leading order in the long-range
Coulomb interaction electrons and holes
are separately conserved, so there is no screening by virtual
electron-hole pairs, and exciton formation destabilizes the putative
gapless state\cite{halperin}.

Using the RG, we can evaluate the anomalous dimension of any physical
operator.  By charge conservation, $[\psi^{\dagger}\psi]=d$.  Due to
the isotropy of the fixed point, there are only two non-trivial values
for the other charge-conserving fermion bilinears. 
We obtain , $[\psi^{\dagger} \Gamma_a \psi] =d+
\eta_1$ , $[\psi^{\dagger} \Gamma_{ab} \psi]=d+\eta_{12} $, , where
$\Gamma_a$ are the (time-reversal invariant) Dirac gamma matrices,
$\Gamma_{ab} = -\frac{i}{2}[\Gamma_a,\Gamma_b]$ are time-reversal odd,
and $a,b=1,2,\cdots,5$.  Using the standard operator insertion
technique, we find $\eta_{1} = - \frac{6 }{15 N_f +4}\varepsilon$ and
$\eta_{12} =- \frac{3 }{15 N_f +4}\varepsilon$.  These operators
describe many physical observables, {\it e.g.} the ``angular momentum" operator $J_z \sim
\psi^{\dagger} (-\Gamma_{34} -\frac{1}{2} \Gamma_{12}) \psi $. 
The negative anomalous dimension of these operators
suggests the a schematic picture of power-law excitons due to
electron-hole attraction.  
For pairing channels, we find positive anomalous dimensions,
consistent with this view.
The local pairing channel has $\eta_{pairing} = \frac{u^*}{5}= \frac{3}{30 N_f + 8}\epsilon$. 

Using these results, we obtain thermodynamic responses such as the
specific heat, $c_v \sim T^{d/z}\approx T^{1.7}$ and the spin
susceptibility $\chi(T) \sim a+b \,T^{(d-z+2\eta_{12})/z}\approx a + b \,
T^{0.5}$, with some constants $a,b$.  Interestingly, the non-linear
susceptibility $\chi_3 = \left.\partial^3 M/\partial H^3\right|_{H=0}
\sim T^{-(3z-4\eta_{12}-d)/z} \approx T^{-1.7}$ diverges, as in spin
glasses but with completely different physics.  Comparing the scaling of
current and electric field gives the usual result $[\sigma_{ij}] =d-2$.  Consequently, the temperature and frequence
dependence of the conductivity is $\sigma(\omega,T) \sim T^{1/z}
\mathcal{F}(\omega/T)$, and a {\em clean, undoped} LAB is therefore a
power-law insulator.  

We now turn to the effect of applied strain and Zeeman field upon the
LAB.  These perturbations break cubic/time-reversal
symmetries, and thus destabilize the LAB.   Due to the isotropic
nature of the LAB fixed point, the response to the Zeeman field alone is
to leading order independent of its direction (the cubic mass $1/M_c$
can be ``dangerously irrelevant'', however -- see below), so we take
it to lie along the (001) direction.  We consider for simplicity tetragonal strain
which preserves $C_4$ rotation about this axis (in the absence of
Zeeman field, the direction of strain is again unimportant).  This
leads to the perturbations
\begin{eqnarray}
{\mathcal H}' = -\delta (J_z^2-\frac{5}{4}) -H (\cos(\theta) J_z + \sin(\theta) J_z^3) ,\label{pert}
\end{eqnarray}
where $\delta$ parametrizes the strain, $h$ is the Zeeman field, and
$\theta$ controls the strength of the cubic Zeeman term allowed by the
cubic symmetry \cite{kane_fu,ewald}. Using the RG results, the dimensions of
these perturbations are $[\delta]=z-\eta_1 \approx 2.1$ and
$[H]=z-\eta_{12}\approx 1.9$; {\it i.e.} strain is slightly enhanced
while Zeeman field is slightly suppressed by interactions.  However,
both dimensions are positive and close to $2$, so that they are
strongly {\em relevant}.  They flow to strong coupling under the RG,
and the fate of the system must be re-analyzed in the limit.

To do so, we assume, and check self-consistently, that interactions
have weak effects at strong coupling, and simply solve the quadratic
Hamiltonian (with $m/M_0=m/M_c=0$) in the presence of the renormalized
$\mathcal{H}'$.  The result depends upon the dimensionless quantities
$\theta$ and the renormalized coupling ratio $\Delta = (\delta/H)_R \sim
\delta/H^{(z-\eta_1)/(z-\eta_{12})}$.  For
$H=0$ ($\Delta=\infty$), we have time-reversal invariance, and we recover the
known result that strain $\delta>0$ induces a gapped, 3d TI
phase, as observed in HgTe \cite{zhang}.  The situation in applied Zeeman
field is more interesting.  Notice that
for $\vec{k} = k\hat{z}$, $J_z$ is a good quantum number, and there is
no level repulsion between bands of different $J_z$.  This allows
(non-degenerate) bands to cross along this axis, which indeed occurs when
$|\Delta|$ is not too large.  Further analysis in
the Supplementary Material shows that these crossings correspond to a
pair of {\em double Weyl points}, with linear dispersion along the $z$
axis and quadratic dispersion normal to it.  These points are strength
$\pm 2$ monopoles in momentum space.  Away from the $k_z$ axis, electron and hole
pockets may accidentally cross the Fermi energy.  If this does not occur, one has a
pristine double Weyl semimetal, which occurs for the angular range
$\theta_1 \le \theta \le \theta_2$, where $\theta_1 =
-\tan^{-1}(\frac{8+4 \sqrt{3}}{7 \sqrt{3} +26})$ and $\theta_2 =
\tan^{-1}(\frac{8-4 \sqrt{3}}{7 \sqrt{3} -26})$ for $\delta=0$, as shown in horizontal axis of 
Fig.~\ref{fig1}.  When $0<|\Delta|<\infty$, we observe
insulating, double Weyl semimetal, and Weyl metal (with coexisting
electron-hole pockets) phases, as shown in Figure~\ref{fig1}.  Note that in
all these phases, the Coulomb interactions become either unimportant
(in the insulator), screened (in the metal),  or marginally irrelevant
(in the Weyl semimetal), justifying our treatment of the phase diagram
to a first approximation.  

More subtle effects may make small modifications to this picture.
Coulomb interactions can destabilize some of the quantum phase
transitions in Figure~\ref{fig1}, leading to intermediate phases.  When the
magnetic field is applied along a low symmetry axis, the double Weyl
points can split into multiple single Weyl points, once the effects of
the cubic mass $1/M_c$ is included, which is dangerously irrelevant in this case.

A striking experimental consequence of this phase diagram is the
AHE due to the Zeeman field, which could originate
either from an external magnetic field or as an exchange field due to
local moments in the material.  The latter is particularly interesting
in light of the experimental results on Pr$_2$Ir$_2$O$_7$, which shows
a large AHE in a regime where the magnetization $M$ is immeasurably
small \cite{nakatsuji,machida}.  On symmetry grounds, $\sigma_{xy}
\neq 0$ {\em implies} $M_z \neq 0$, but evidently $\sigma_{xy}$ is
unusually large relative to $M$.  This behavior is in fact
characteristic of the LAB:  since the
Hall conductivity has dimensions of inverse length, we expect
$\sigma_{xy} \sim H^{1/(z-\eta_{12})} \approx H^{0.51}$.  In the
situation relevant for Pr$_2$Ir$_2$O$_7$, the Zeeman field is generated by
Kondo exchange with the Pr moments, so $H \sim J_K M$, with $M$ 
the (dimensionless) Pr magnetization, which implies a highly
unconventional sublinear dependence of $\sigma_{xy}$ on $M$ for the
pristine LAB.  If the Fermi level is displaced
from the band touching by an amount $\epsilon_F$, then we expect,
treating the above power as a square root, $\sigma_{xy} \sim
\frac{e^2}{h} \sqrt{\frac{m H}{\hbar^2}} \mathcal{S}(H/\epsilon_F)$, where
$\mathcal{S}$ is a scaling function (see Supp. Mat.).  This gives an order of magnitude
quantitative estimate
\begin{equation}
  \label{eq:8}
  \sigma_{xy} \sim 10^3 \Omega^{-1} cm^{-1} \times
  \left\{ \begin{array}{cc} \sqrt{\frac{m}{m_e}}  \sqrt{\frac{J_K}{\rm Ry}}
      \sqrt{M}, & J_K M \gg \epsilon_F \\
      \sqrt{\frac{m}{m_e}} \frac{J_K}{\sqrt{\epsilon_F {\rm Ry}}} M,
      &  J_K M \ll \epsilon_F \end{array}\right. \; ,
\end{equation}
where Ry=13.6eV is the Rydberg, and $m_e$ is the electron mass.  The
lower regime gives $\sigma_{xy} \sim 0.1 \Omega^{-1} cm^{-1}$,  within
an order of magnitude of observations 
in Pr$_2$Ir$_2$O$_7$,\cite{machida} for parameters $m \sim 20m_e$ (estimated from
the calculations in Ref.~\onlinecite{ybkim2}) , $\epsilon_F \sim $10meV,
$J_K \sim$100K (estimated from the measured Curie-Weiss temperature),
and $M \sim 0.01$.

Another interesting experimental observation in Pr$_2$Ir$_2$O$_7$ is a
diverging magnetic Gruneisen number $\Gamma_H = \frac{1}{T}
\left. \frac{\partial T}{\partial H}\right|_S$ in the zero field
limit \cite{si, gruneisen, 227}.  In a purely electronic system with no local moment
contribution to the entropy, we can readily obtain the
behavior of the LAB in the low temperature limit,
\begin{eqnarray}
\Gamma_H(H,T) = - \frac{d- y_0 z}{y_0 (z+\eta_H)} \frac{1}{H} ,\label{eq:3}
\end{eqnarray}
which depends upon the exponent $y_0$ defining the temperature dependence of the specific heat, $C \sim T^{y_0}$, 
of the LAB phase.  In the Weyl metal, $y_0=1$, and $\Gamma_H<0$,
while for isolated double Weyl points, $y_0=2$, and $\Gamma_H>0$.
Thus one may imagine a sign change of the Gruneisen number when field
or strain is varied.  Note that in Pr$_2$Ir$_2$O$_7$, there is
certainly a large local moment contribution to the entropy, so that
  Eq.~\eqref{eq:3} is not literally applicable.  Nevertheless, the LAB
  physics may play some role in this quantity.  

In conclusion, we have described a novel NFL phase occurring in
correlated strong SOC systems, with a natural connection to the
pyrochlore iridates.  Even with weak correlation effects, some of the
phenomena discussed here can be observed with only minor
modifications, and it would be interesting to search for them in HgTe.
 Future theoretical studies should include more comprehensive
 treatment of breaking of cubic symmetries, and the effects of
 disorder and doping.  

We thank T. Hsieh, L. Fu, and P. Gegenwart for discussions on Luttinger model
and Gruneisen number. This work was supported by NSF Grant DMR-1151208 
and Hellman Family Foundation (EGM, CX), 
the David and Lucile Packard Foundation, the Alfred P. Sloan Foundation  (CX), 
NSF--DMR--1206809 (LB), and NSERC, CIFAR (YBK).
YBK and LB acknowledge the support and hospitality from the Aspen
Center for Physics, funded by NSF grant PHY-1066293, and the KITP,
funded by NSF grant PHY--1125915.

\section{Supplementary material}

\subsection{Representations of the Hamiltonian}

In this section, we provide more information about the different
representations used in the main text.  The Hamiltonian reads
\begin{eqnarray}
\mathcal{H}_0(k)&=&\alpha_1 k^2 + \alpha_2 (\vec{k} \cdot \vec{J})^2 + \alpha_3 (k_x^2 J_x^2 + k_y^2 J_y^2 +k_z^2 J_z^2) \nonumber \\
&=& \frac{k^2}{2 \tilde{M}_0} + \frac{ (\frac{5}{4}k^2-\vec{k} \cdot \vec{J})^2}{2m} - \frac{ (k_x^2 J_x^2 + k_y^2 J_y^2 +k_z^2 J_z^2)}{2 M_c} \nonumber \\
&=&\frac{d_a(k)}{2m} \Gamma^a + \frac{k^2}{2M_0} + \frac{d_4(k) \Gamma^4 + d_5(k) \Gamma^5}{2M_c} .
\end{eqnarray}
The first line uses the conventional Luttinger parameters
($\alpha_{1,2,3}$) in the $j=3/2$ matrix representation, and the
second line is the form used in the main text.  For the purpose of
computations, it is convenient to introduce the Clifford gamma
matrices ($\Gamma_{a}$) in the third line as in the paper by Murakami
{\it et al.} \cite{murakami}.
\begin{eqnarray}
&&\epsilon_a(k)= \frac{d_a(k)}{2m} , \nonumber \\
&& d_1(k)= -\sqrt{3}k_y k_z \, , \,\,  d_2(k)= -\sqrt{3}k_x k_z \,, \,\, d_3(k)= -\sqrt{3}k_x k_y  , \nonumber \\
&& d_4(k) = \frac{-\sqrt{3}}{2}(k_x^2 -k_y^2) \,, \,\, d_5(k) =
\frac{-1}{2}(2 k_z^2 - k_x^2 -k_y^2) .
\end{eqnarray} 
It is straightforward to relate the masses used in the main text and
the Luttinger $\alpha_i$ parameters. This can be done by expressing
the spin operators in terms of gamma matrices, using the equalities
\begin{eqnarray}
&& J_x = \frac{\sqrt{3}}{2} \Gamma_{15} - \frac{1}{2} (\Gamma_{23} - \Gamma_{14}) \ ,\nonumber \\
&& J_y = -\frac{\sqrt{3}}{2} \Gamma_{25} + \frac{1}{2} (\Gamma_{13} + \Gamma_{24}) \ ,\nonumber \\
&& J_z = - \Gamma_{34} - \frac{1}{2} \Gamma_{12} \ ,
\end{eqnarray}
where $\Gamma_{ab} = \frac{1}{2 i} [\Gamma_a, \Gamma_b]$ is used.

\subsection{Weyl semimetal}

Here we consider how the Weyl semimetal appears, taking for simplicity
the case $\delta =0 , \theta_1 < \theta < \theta_2$.  It is
straightforward to generalize this to the cases with $\delta\neq 0$.
For convenience, we set $m=1/2$.
For $\vec{k}=(0,0,k_z)$, the Hamiltonian in the presence of the Zeeman
field, $h$ (along the same axis), becomes
\begin{eqnarray}
\mathcal{H}(k_z \hat{z})&=& k_z^2 (\frac{5}{4}-J_z^2)-H ( \cos(\theta) J_z +\sin(\theta) J_z^3). \nonumber 
\end{eqnarray}
Clearly, the energy eigenstates can be labeled by both $k_z$ and
$J_z=\pm 1/2, \pm 3/2$. One can readily see that level crossings occur
between the two bands with $J_z=1/2$ and $J_z=-3/2$.  In the vicinity
of these crossing points, the other states with $J_z=-1/2, +3/2$ can
be discarded, and the Hamiltonian projected onto the two level
subspace of low energy states.   We introduce Pauli matrices in this
subspace, so that $\tau^z =
|\tfrac{1}{2}\rangle\langle\tfrac{1}{2}| -
|-\tfrac{3}{2}\rangle\langle-\tfrac{3}{2}|$, $\tau^+ = (\tau^x + i
\tau^y)/2 = |\tfrac{1}{2}\rangle\langle-\tfrac{3}{2}| =
(\tau^-)^\dagger$, and  $\tau^0 =
|\tfrac{1}{2}\rangle\langle\tfrac{1}{2}| +
|-\tfrac{3}{2}\rangle\langle-\tfrac{3}{2}|$, which is the identity
matrix in the $2\times 2$ subspace. We define two energy
parameters, $\epsilon_{-3/2} = H (\frac{3}{2}\cos(\theta) +
\frac{27}{8} \sin(\theta))$ and $\epsilon_{1/2} =-H
(\frac{1}{2}\cos(\theta) + \frac{1}{8} \sin(\theta)) $.  Then the
reduced Hamiltonian becomes
\begin{eqnarray}
&&\mathcal{H}_{2} = \epsilon_{+} \tau^0+ (\epsilon_{-}+d_5(k)) \tau^z +d_4(k) \tau^x - d_3(k) \tau^y, \nonumber \\
&&\epsilon_{\pm}=(\epsilon_{-3/2}\pm\epsilon_{1/2})/2.
\end{eqnarray}
There are level crossing points at $k_x=k_y=0$ and $k_z=\pm
K$, with $K=\sqrt{\epsilon_-}$.  We expand around these points, letting $k_x=p_x$,
$k_y=p_y$ and $k_z =\pm
K+ p_z$, which gives
$\mathcal{H}_2(\pm K\hat{z}+\vec{p}) = \epsilon_+ \tau^0 + 
\mathcal{H}_{2}^\pm(\vec{p})$, with, to leading order in $\vec{p}$,
\begin{eqnarray}
\mathcal{H}_{2}^\pm = \mp v\, p_z \tau^z +d_4(p) \tau^x + d_3(p) \tau^y \ , \nonumber
\end{eqnarray}
with $v=2\sqrt{\epsilon_-}$.  The energy spectrum is
\begin{eqnarray}
E(p)=\pm \sqrt{v^2 p_z^2 + \frac{3}{4}(p_x^2+p_y^2)^2}. \nonumber
\end{eqnarray}
We see that the electrons disperse linearly along the field and
quadratically orthogonal to it, near the touching point. This can be understood as a
consequence of 4-fold rotational symmetry around the $z$ axis.  Since
$|J_z\rangle \rightarrow e^{i \pi J_z/2} |J_z\rangle$ under such a
rotation, the operators $\tau^\pm$ carry a net angular momentum of
$\pm 2$, and therefore must couple to the ``d-wave'' combinations of
$p_x$ and $p_y$, which are precisely given by $d_3(p)$ and $d_4(p)$.  

Though the quadratic dispersion normal to the field is due to
symmetry, the touching itself has a topological character.  To see it,
it is convenient to define the reduced Hamiltonian in the form
\begin{equation}
  \label{eq:4}
  \mathcal{H}_{2}^\pm = \vec{b}_\pm(\vec{p}) \cdot \vec\tau,
\end{equation}
with
\begin{equation}
  \label{eq:5}
  \vec{b}_\pm(\vec{p}) = \left( \begin{array}{c} -\frac{\sqrt{3}}{2}(p_x^2-p_y^2) \\
    \sqrt{3} p_x p_y \\ \mp v p_z \end{array}\right) .
\end{equation}
From this, one can define the $U(1)$
Berry flux, which is analogous to a magnetic field in momentum space,
\begin{equation}
  \label{eq:6}
  \mathcal{B}_\pm^\mu = \frac{1}{8\pi}\epsilon^{\mu\nu\lambda} \hat{b}_\pm
  \cdot \partial_\nu \hat{b}_\pm \times \partial_\nu \hat{b}_\pm,
\end{equation}
where $\hat{b}_\pm = \vec{b}_\pm/|\vec{b}_\pm|$, and the derivatives are with
respect to $\vec{p}$.  By construction, the magnetic field is
divergenceless, $\partial_\mu \mathcal{B}_\pm^\mu=0$, away from points of
singularity where $\vec{b}_\pm$ vanishes.  However, the band touching
point is such a singularity, and it is in fact a source of Berry flux.  To
see this, one may compute the integral of the flux of
$\mathcal{B}^\mu$ through a sphere around one of the touching points.
It is straightforward to compute this integral, and by doing so one
finds
\begin{equation}
  \label{eq:7}
  \partial_\mu \mathcal{B}^\mu(\vec{p}) = \pm 2 \delta(\vec{p}).
\end{equation}
Thus each band touching is a source of {\em two} quanta of Berry flux,
and can be therefore considered a {\em double Weyl point}.  The net
Berry flux of both double Weyl points added together vanishes, which
is required as the Brillouin zone is a closed manifold without any
boundary through which a net flux may escape.  

The Berry flux of each double Weyl point is directly related to the
Hall conductivity.  Indeed, one can show quite generally that the Hall
conductivity is proportional to separation of the points,\cite{anton}
\begin{eqnarray}   
\sigma_{xy} = &=&2\times \frac{e^2 }{h} \times \frac{2K}{2\pi} \ , \nonumber 
\end{eqnarray}
where the first factor of $2$ is due to the doubled nature of the Weyl
points.  This is valid at zero temperature when the Fermi level is at
the energy of the double Weyl points.  Neglecting the small
corrections to the exponent due to Coulomb interactions, we have $K
\sim \sqrt{m H}$, which explains the estimate in Eq.~\eqref{eq:8} of
the main text.
 
\subsection{RG equations}

Here we describe the renormalization of the fermionic propagator,
which was not presented in the main text.  We start with the
Hamiltonian
\begin{eqnarray}
\mathcal{H}_0(k)&=& \frac{k^2}{2M_0} \Gamma^0+\epsilon_a(k) \Gamma^a  + \frac{d_4(k) \Gamma^4 + d_5(k) \Gamma^5}{2M_c} \ , \nonumber
\end{eqnarray}
where we defined $\Gamma^0$ as the $4\times 4$ unit matrix to emphasize the matrix structure. 
The corresponding effective Lagrange density is
\begin{eqnarray}
\mathcal{L}_{\rm eff} = \psi^{\dagger} (-i \omega_n  \Gamma^0
+\mathcal{H}_0(k) - \Sigma_{f}(k,i \omega_n))\psi \nonumber .
\end{eqnarray}
Here we have included the one-loop self energy needed for the
$\varepsilon$ expansion. It is given by 
\begin{eqnarray}
  && \Sigma_{\psi}(k, i \omega_n) = \\ 
  && -g^2 \int_q T \sum_{\Omega_n} G_f(k+q, i \omega_n+i \Omega_n) G_{\varphi} (q,i \Omega_n) \ ,\nonumber 
\end{eqnarray}
where $G_f^{-1}(k,i \omega) = -i \omega \Gamma^0 +\mathcal{H}_0(k)$,
$G_{\varphi}^{-1}(q,i \Omega) = c_0 q^2$ are used.  Note that the
bosonic propagator does not have frequency dependence since it
represents the instantaneous Coulomb interaction.  By integrating out
high momentum degrees of freedom, we investigate the renormalization
of fermion parameters.

Frequency and momentum dependences of the self energy determine
anomalous dimensions of the fermion field and mass terms,
respectively.  At one-loop level, the fermion self energy does not
have frequency dependence due to the instantaneous Coulomb
interaction, so no anomalous dimension of the fermion field appears.
Moreover, the frequency-independence makes one mass coupling ($1/M_0$)
unchanged at one loop level and it is because frequency and this mass
term have the same unity matrix structure.  Note that both corrections
naturally appear in higher-loop contributions.

On the other hand, non-trivial renormalizations of isotropic mass ($m$) and anisotropic mass ($M_c$) appear even at one-loop level.
For example, the renormalization of isotropic mass term can be read off by calculating the third component ($\Gamma^3$) of the self energy, 
\begin{eqnarray}
&&\gamma^3(k) \equiv {\rm Tr}\left(\Gamma^3 \frac{\partial^2}{\partial k_x \partial k_y} \Sigma_{\psi}(k, 0) \right)  / {\rm Tr} (\Gamma^3 \Gamma^3) \nonumber \\
&=& -\frac{g^2}{c_0}\frac{\partial}{\partial k_x \partial k_y} \int_q \frac{\epsilon_3(k+q)}{ \sqrt{\left(\frac{(k+q)^2}{2m}\right)^2+ \frac{m+2 M_c}{4 m M_c^2} p_c(k+q) }} \frac{1}{q^2}. \nonumber 
\end{eqnarray}
It can be understood as the self energy corrrection to the original Hamiltonian,
\begin{eqnarray}
\epsilon_3(k) &\rightarrow& \epsilon_3(k)- \gamma^3(0) k_x k_y 
\nonumber \\
&=&\epsilon_3(k)\left(1+ u f_1(\frac{m}{M_c}) \log(\frac{\Lambda}{\mu}) \right).
\end{eqnarray}
In the last line, the momentum integration is done for $\mu < q < \Lambda$.
The correction to the isotropic mass term, $\delta (1/m) = \frac{u}{m}  f_1(\frac{m}{M_c}) \log(\frac{\Lambda}{\mu})$ can be regularized in the standard way.
Along the same line, one can determine the renormalization of the anisotropic mass term by calculating the fourth component ($\Gamma^4$) of the self-energy.

\begin{figure}
\includegraphics[width=3.0in]{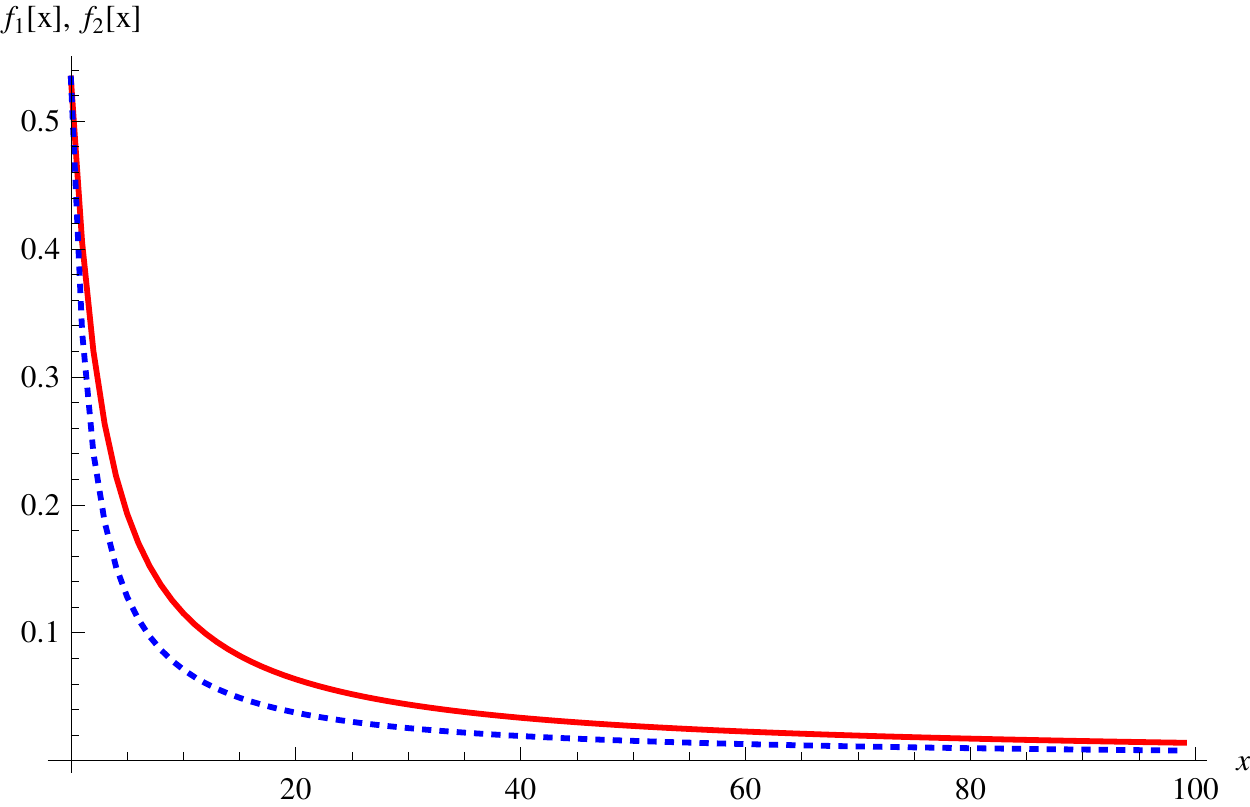}
\caption{ Numerical evaluations of the renormalization functions, $f_1(x), f_2(x)$. The solid (red) and dotted (blue) lines are for $f_1(x)$ and $f_2(x)$ ($f_1(0)=f_2(0) = 8/15$).
One can show that $f_1(x)-f_2(x)$ decays as $1/x$ in the large $x$ limit.
} \label{f12}
\end{figure}

\begin{figure}
\includegraphics[width=3.0in]{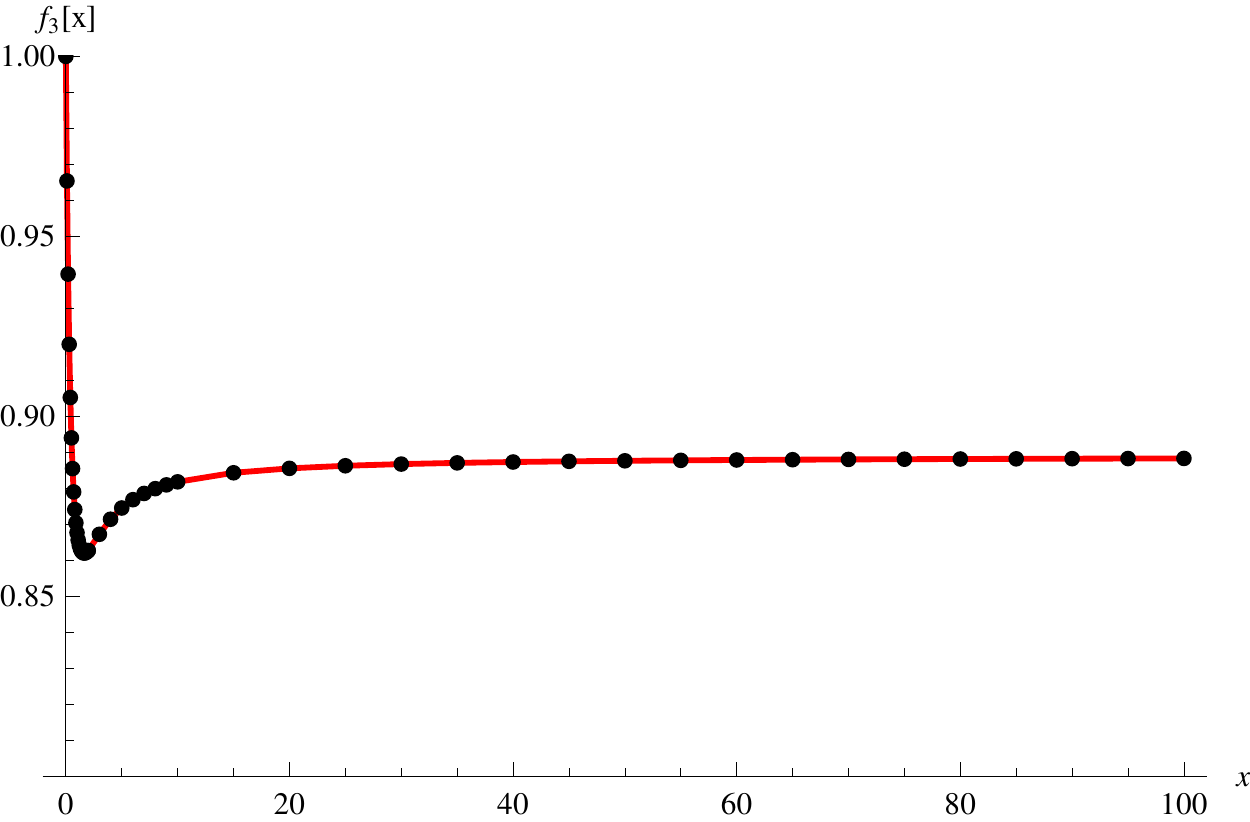}
\caption{ Numerical evaluations of the universal function, $f_3(x)$. 
One can see that $f_3(x)$ becomes saturated as $x$ becomes large.
} \label{f3}
\end{figure}

These considerations lead to the full RG equations as follows.
\begin{eqnarray}
&&\frac{d}{d l} (\frac{1}{m}) =\left(z-2 + u  f_1(\frac{m}{M_c})\right) (\frac{1}{m}) \ , \nonumber \\
&& \frac{d}{d l}(\frac{m}{M_0}) = -u f_1(\frac{m}{M_c}) \frac{m}{M_0} \ , \nonumber \\
&& \frac{d}{d l}(\frac{m}{M_c}) = -u F_2(\frac{m}{M_c}) \ , \nonumber \\
&&\frac{d }{d l} (u) = \varepsilon \, u - N_f F_1(\frac{m}{M_c}) u^2 \ . \nonumber  \label{beta}
\end{eqnarray}
Here the following functions of anisotropy are introduced. 
\begin{eqnarray}
&&f_1 (x) = \frac{2}{ \pi} \int d \Omega_{k} \frac{\hat{k}_x^2\hat{k}_y^2}{\sqrt{1+x(2+x) p_c(\hat{k})} } \ , \nonumber \\
&&f_2 (x) = \frac{1}{2 \pi}  \int d \Omega_k\frac{(\hat{k}_x^2-\hat{k}_y^2)^2 }{\sqrt{1+x(2+x)p_c(\hat{k})} } \ ,\nonumber \\
&&f_3 (x) = \frac{3}{8 \pi} \int d \Omega_k   \frac{x^4 k_x^2 (\hat{k}_y^2-\hat{k}_z^2)^2 }{\Big(\sqrt{1+x(2+x)p_c(\hat{k})}\Big)^5} + O(\frac{1}{x}) \ , \nonumber \\
&&F_1 (x) = 2 f_3 (x)+  \frac{1}{N_f} f_1 (x) \ , \nonumber \\
&&F_2 (x)=   ( 1+x)\left(f_1 (x) -f_2 (x)\right) \ , \nonumber 
\end{eqnarray}
where $p_c(k)  = \sum_{i=1}^3 k_i^4 -\sum_{i \neq j} k_i^2 k_j^2$ is used. 
The first two functions, $f_1$ and $f_2$, are from the fermion self energy and the third function, $f_3$, is from the boson self-energy calculation. 
The asymptotic form of $f_3$ is shown above while its complete form can be obtained by evaluating the 
bosonic self energy function directly. 
Numerical evaluations of these functions are illustrated in Figs. \ref{f12} and \ref{f3}.
Useful numerical values are 
\begin{eqnarray}
&&N_f F_1(0)= \frac{30 N_f +8}{15} \ , \quad F_2(0) = 0 \ , \quad F_2^{'} (0) \sim 0.152 \ , \nonumber \\
&&f_1(0)=f_2(0)=\frac{8}{15} \ , \quad f_3(0) = 1\ , \nonumber \\
&&\lim_{x \rightarrow \infty} f_1(x) =\lim_{x \rightarrow \infty} f_2(x) =0 \ , \nonumber \\
&&\lim_{x \rightarrow \infty} f_3(x) \sim 0.888. \nonumber
\end{eqnarray}

We note that there is one subtle issue in the $\varepsilon$ expansion 
for regularization of the loop corrections. 
Since the gamma matrices are defined in specific dimensions, 
there is some ambiguity for the choice of gamma matrices in dimensional regularization.
There have been some suggestions as to how to apply the $\epsilon$ expansion to Gamma matrices. \cite{ab2}

In this letter, we use one of possible regularization schemes;
the angle dependence is first integrated in the original dimension and then dimensional regularization is
used for the integrals over the magnitude of momentum.  
One reason for using this scheme is our need to keep track of the cubic anisotropy, which is specifically 
defined in three spatial dimensions. 
Formally, the momentum shell integration gives $
\int_{\Lambda e^{-dl}}^{\Lambda} \frac{d^d q}{(2 \pi)^d} \frac{1}{q^4} \rightarrow \frac{d l}{8 \pi^2}
$.
Note that different regularization schemes only change numerical coefficients and do not change important physics.


\begin{thebibliography}{99}
\bibitem{large_n} In this paper, we perform the $\varepsilon=4-d$ expansion using the $4 \times 4$ gamma matrices. 
Alternatively, one could use $8 \times 8$ matrices in $d=4$ with a half of the number of fermion flavors and 
calculate logarithmic divergences directly.
\end{thebibliography}
\end{document}